# Cognitive factor forming an individual constituent in a driver model inferred from multiplicatory relationships between cognitive sub-factors

Firas Lethaus*[1,2], Robert Kaul[3]

[1]*Volkswagen AG, Brieffach 011/80830, 38436 Wolfsburg, Germany*
[2]*XAI Lab, Alberta Machine Intelligence Institute (AMII), Department of Computing Science, University of Alberta, Edmonton, Alberta, Canada T6G 2E8*
[3]*Independent Senior Research Scientist*

*Corresponding author:
Firas Lethaus, Volkswagen AG, Brieffach 011/80830, 38436 Wolfsburg, Germany
Email: firas.lethaus@gmail.com

Cognitive factor forming an individual constituent in a driver model inferred from multiplicatory relationships between cognitive sub-factors

Abstract

In order to reproduce human behaviour in dynamic traffic situations, a computational representation of the requisite mental processes used to carry out the complex driving tasks is required. A single cognitive factor has been developed and forms a crucial component in our driver model. This cognitive factor is composed of cognitive sub-factors: distraction, anticipation, stress, and strain. It has been defined such that these sub-factors have a multiplicatory relationship with one another. In addition, each of these sub-factors comprises the visual, auditory, tactile-kinetic, and verbal information channel individually. This results in a lattice-like network of relationships, which can be expanded modularly in horizontal and vertical directions. The conceptual topology and the characteristics of this multiplicatory, operation-based cognitive factor are presented.

Keywords: Cognitive factor, Driver model, Multiplicatory relationships, Advanced driver assistance systems, Adaptive automation, Computational representation.

# 1 Introduction

The act of negotiating traffic in a vehicle is a very complex monitoring and control task requiring continuous information processing. This task is very demanding for the driver as it requires targeted sensomotoric actions to be carried out perennially. The command variable, that is, the output of the driving task, is determined by the pathway of the road, other vehicles and road users as well as by environmental conditions and runs as a continuous setpoint / actual value comparison within the driver-vehicle-road system (Bubb, 2001). It is therefore referred to as vehicle guidance, if a vehicle is moved from the place of departure to its target location while using sensory information and an on-board motoric input interface (Bubb, 2001; Donges, 2009). In order to better understand the operational stages when controlling a vehicle, the actual act of driving itself is abstracted and can schematically be explained via a model. According to Kramer (2008), most human controller models are subject to the problem of 'identifiability of model parameters', which can only be tackled via explicit modelling of the kinematics of driving.

Automation can be an advantage or disadvantage to human performance, and usually becomes noticeable when it fails in operation or does not accomplish the desired outcome (Fitts, 1951; Wickens & Kessel, 1979; Wickens & Kessel, 1980; Wiener & Curry, 1980; Rasmussen & Rouse, 1981; Wickens & Kessel, 1981; Wickens et al., 1998; Sheridan, 2002). Adaptive automation (AA) poses a design philosophy, where functions of a human-machine system (HMS) are dynamically divided between human and machine agents. This is accompanied by an adaptation to the operator's current situation and current state. Here, the overall performance and safety of an HMS is improved by the machine's support and take-over of tasks, which prevent the human agent from exhibiting slumps in performance. With an adaptive system, the distribution of a task is controlled by the machine agent. In order to distribute a task to an optimum level, a sufficient knowledge base is required pertaining to the requirements of the operators when carrying out their tasks. In the same manner, the system needs to monitor the state of the operator and the situation; so that it would be able to respond to any changes which may occur. AA provides an opportunity to make Advanced Driver Assistance Systems (ADAS) adaptive to traffic and driver state demands, such as workload, distraction, anticipation, expectancy, or intention. Any planned actions coming from the driver can be considered by an ADAS as a means to enhance a general information strategy, which forms an assistance system's central control unit. The ability to model the mental processes behind the execution of complex driving tasks allows for the development of ADAS to be accelerated, also by reducing the number of time-consuming and expensive studies required as human behaviour in dynamic traffic situations can be reproduced. Additionally, it permits holistic assistance strategies containing knowledge about driver behaviour to be amended (Rasmussen & Rouse, 1981; Sheridan, 2002; Vasconez et al., 2020; Wickens & Kessel, 1979; Wickens & Kessel, 1980; Wickens & Kessel, 1981; Wickens et al., 1998; Wiener & Curry, 1980).

When it comes to mental processes and their cognitive interference, executive functions (EF) need to be considered. EF denote top-down mental processes that are required when having to focus or to pay attention, which involves effort as it is easier to hand over to automation than to think about the next step of action (Burgess & Simons 2005; Diamond 2013; Gavas et al., 2018). According to Lehto et al. (2003), there are three core EF: *inhibitory control, working memory*, and *cognitive flexibility*. Based on these EF, derivatives or higher order EF can be formed such as planning, reasoning, and problem solving (Collins & Koechlin 2012; Lunt et al., 2012). *Inhibitory control* refers to being able to control one's attention, behaviour, thoughts, or emotions in order to override strong internal biases or external enticement. It further allows for selective attention as well as the suppression of attention with respect to disturbing stimuli to take place, which poses interference control at perception level. Salient stimuli, whether they are visual, auditory, or tactile-kinetic, draw involuntary attention and are processed bottom-up, whereas we are also able to voluntarily ignore certain stimuli and, thus, execute attentional control where information is processed top-down (Diamond 2013; Posner & DiGirolamo, 1998). *Working memory* denotes holding and processing information that is no longer present at perception level, and is crucial for establishing meaning over time, grasping information of the past that is linked with information to come. This forms the basis of our ability to link isolated elements and to split up things that form a complete unit, which is the prerequisite of applying conceptual knowledge. Working memory is distinguished from short-term memory, in that short-term memory refers to holding information only and working memory refers to holding and manipulating information (Baddeley & Hitch, 1994; Diamond, 2013). *Cognitive flexibility* represents the ability to take different perspectives on things or opinions. This requires a previous perspective to be inhibited and a different perspective to be loaded into the working memory, which makes *inhibitory control* and *working memory* serve as the pillars of cognitive flexibility (Diamond, 2013).

Conscious involvement in a task over a longer period of time is usually accompanied by allowing our attentional focus to deviate. This dynamic nature of conscious involvement is referred to as *mind-wandering*. Here, attention shifts from the current task to non-related thoughts and emotional states. It is assumed that the information-processing demands of mind-wandering are linked to perceptual decoupling in order to defy constraints of the moment. It is further assumed that the mind-wandering state's content arises from episodic and affective processes and that its regulation depends on executive control (Smallwood & Schooler, 2015).

The production of a suitable cognitive driver model entails describing these mental processes such that the key characteristics of the driver behaviour being considered are captured. The description needs to comprise constituents which allow the behaviour to be expressed formally, eventually resulting in a transfer from driver behaviour to computational simulations of behaviour. In order to achieve this goal, the first task is to identify and model the cognitive parameters, which are pertinent when describing driver behaviour. As a result, a single Cognitive Factor (CF), which is composed of cognitive sub-factors (CSF), has been developed and forms an essential constituent within our driver model. The conceptual topology and the characteristics of this multiplicatory, operation-based CF are presented.

## 2 Model Topology and Characteristics

The CF is composed of the CSF *distraction* (D), *anticipation* (An), *stress* (St), and *strain* (Sn). However, in order to qualify as valid CSF, it is important that they are sufficiently separate and independent from one another theoretically.

*Distraction* can be defined as a process of deflecting the attention of an individual or group from a desired area of focus and thereby reducing the reception of desired information. This may be caused by an inability to pay attention or an indifference towards the object of attention. There are external and internal sources of distraction, both of which interfere with focus. External sources of distraction comprise features such as visual or acoustic cues as well as social interactions. Internal sources include fatigue, hunger/thirst, distress, and grief (Schießl, 2008a; Schießl, 2008b; Bingham, 2014). Here, distraction is expressed by a positive real number in the range

$$[1,2] \; ,$$

as this may represent decelerated behaviour leading to slower responses, that is, longer reaction times. A value equal to 1 would express no effect.

*Anticipation* refers to the assumption that an event will take place in the future based on previous experience. In mechanistic terms, this can be expressed as a mechanism having the nature of an anticipatory partial goal response enabling goal-oriented chains of action (Spence, 1956; Hull, 1952). The subjective probability linked to that event may vary (Balkenius, 1995). Here, anticipation is expressed by a positive real number in the range

$$(0,1] \; ,$$

as this may represent vigilant behaviour based on cognitive bias leading to faster responses, that is, shorter reaction times. This may also refer to positive priming effects (Meyer & Schvaneveldt, 1971; Schvaneveldt & Meyer, 1973; Meyer et al., 1975; Kolb & Whishaw, 2009). A value equal to 1 would express no effect.

In terms of work done to distinguish *stress* and *strain* theoretically, definitions have been produced usually discussed in the context of explaining the term 'workload' (Vollrath & Schießl, 2004). According to ISO 10075-1:2017 (2017), psychological *stress* is defined as 'the total of all assessable influence impinging upon a human being from external sources and affecting it mentally'. In contrast, *strain* is defined as 'the immediate effect of mental stress on the individual (not the long-term effect) depending on his/her individual habitual and actual preconditions, including individual coping styles.' Hence, psychological *stress* is considered equal for all human beings, while resulting *strain* is different for an individual due to unpredictable preconditions and coping strategies (Schießl, 2008a; Schießl, 2008b; Bingham, 2014, Andrianov et al., 2021). Here, stress and strain are individually expressed by a positive real number in the range

$$[1,2]$$

as this may represent decelerated behaviour leading to slower responses, that is, longer reaction times. A value equal to 1 would express no effect.

The CF module receives inputs from the information processing module and produces outputs to the response module (see Figure 1). These outputs are the product of all four CSF. Different to distraction, it is assumed that anticipation, stress, and strain are based on previous experience and therefore influenced by the long-term memory (LTM) and working memory (WM) (Baddeley, 2000; Smallwood & Schooler, 2009; Diamond, 2013; Mooneyham & Schooler, 2013; Smallwood & Schooler, 2015).

---------------- Insert figure 1 about here -----------------

The CF has been defined such that, the individual CSF are in a multiplicatory relationship with one another. Figure 2 shows a graphical representation of the CF's conceptual topology, where each of the CSF comprises the visual, auditory, tactile-kinetic, and verbal information channel independently and forms corresponding nodes. This results in a lattice-like network of relationships, which can be expanded modularly in horizontal (refers to information channels) and vertical (refers to CSF) directions. The motoric response is specific to the respective information channel (see Figure 2).

---------------- Insert figure 2 about here -----------------

The CF for a given channel $n$, consists of four CSF related as follows,

$$CF_n = D_n \; An_n \; St_n \; Sn_n \;, \tag{1}$$

where

$$D \neq 0; \; An \neq 0; \; St \neq 0; \; Sn \neq 0 \;. \tag{2}$$

The CSF consist of the visual (V), auditory (A), tactile-kinetic (TK), and verbal (VB) information channels independently, such that

$$CF_V = D_V \; An_V \; St_V \; Sn_V \;, \tag{3}$$
$$CF_A = D_A \; An_A \; St_A \; Sn_A \;, \tag{4}$$
$$CF_{TK} = D_{TK} \; An_{TK} \; St_{TK} \; Sn_{TK} \;, \tag{5}$$
$$CF_{VB} = D_{VB} \; An_{VB} \; St_{VB} \; Sn_{VB} \;, \tag{6}$$

where the CSF can take on values in the following ranges:

$$D = [1, 2] \;, \tag{7}$$
$$An = (0, 1] \;, \tag{8}$$
$$St = [1, 2] \;, \tag{9}$$
$$Sn = [1, 2] \;. \tag{10}$$

The total CF is defined as the product of the CSF, where

$$CF_{total} = CF_V \; CF_A \; CF_{TK} \; CF_{VB} \;. \tag{11}$$

The overall product value $CF_{total}$ may be interpreted as follows within the driver model context:

$$CF_{total} = (0, 1) \quad \text{acceleration effect} \;, \tag{12}$$
$$CF_{total} = 1 \quad \text{no effect} \;, \tag{13}$$
$$CF_{total} = (1, 2) \quad \text{delay effect} \;, \tag{14}$$
$$CF_{total} = [2, 9] \quad \text{extreme delay effect} \;. \tag{15}$$

## 3 Model Output

The numeric value of the resulting $CF_{total}$ can be interpreted as having either an acceleration or delay effect on the transformation of environmental cues into motoric reactions. The CSF are expressed by positive real numbers in the range greater than 0 and less than or equal to 2. Due to the multiplicatory relationship between the CSF, this has the useful effect of allowing CSF with a value of 1 to have no effect within the CF. If three CSF take on a value of 1, the fourth CSF can be considered separately in isolation. If all four CSF take on a value of 1, the CF has no effect within the driver model. In order to numerically express the CSF in the range given, empirical studies will seek to determine the degree of granularity and the standardised scale that can be transferred from qualitative and quantitative research.

In an attempt to give an example that illustrates the topology of the CF and the multiplicatory relationships between CSF (see Figure 2) in a more transparent manner, the CSF nodes and their allocated values have been translated into a heatmap profile showing an overall product value $CF_{total}$, which is displayed in Figure 3. Here, each heatmap cell represents a range of values presented in the legend to the right. In the example given in Figure 3, the overall product value $CF_{total}$ has a value 1.06, which denotes a slight delay effect. In a next step, $CF_{total}$ may serve as a cognitive weight when adding context to a, for instance, response / reaction time (RT) by multiplying it with $CF_{total}$ :

$$[contextual\ RT] = [CF_{total}]\ [RT]\ . \qquad (16)$$

We therefore present a measure called *contextual RT ($RT_C$)*. $RT_C$ may, for example, be utilized, when estimating the expected RT of a driver during the run-up of a hand-over procedure in an automated vehicle, being suitable for automation level 2, 3 , and 4 (SAE International, 2019).

---------------- Insert figure 3 about here -----------------

# 4  Discussion and Outlook

This paper introduces a single CF to be used as a crucial component in a driver model to augment an ADAS or AA system. The topology and the characteristics of the CF and its component CSF have been presented. It remains to be demonstrated empirically, to what degree the numeric range of the cognitive parameters presented as being suitable for CSF maps to the range of the CSF that would be measured in complex traffic situations and eventually to the overall product value $CF_{total}$. In the same way that distinct gaze patterns were found to precede specific driving manoeuvres (Lethaus & Rataj, 2007; Lethaus & Rataj, 2008; Lethaus et al., 2011; Lethaus, 2013; Lethaus et al., 2013; Lethaus et al., 2021), distinct levels of distraction, anticipation, stress, and strain may be found to relate to specific driving situations given specific conditions when carrying out complex driving tasks.

The inclusion of information pertaining to current and future driver behaviour as well as drivers' cognitive state has the potential to enhance AA and ADAS functions with valuable knowledge and they experience a significant additional value, since they can adapt to the drivers' behaviour in real-time. Thus, drivers' intended actions and future cognitive states can be considered in an information strategy comprising ADAS. The development of such functions would require the implementation of a prediction model, linking past and future behaviour, into an online compliant system, which would need to consider other challenges, such as processing a continuous stream of potentially very noisy data. This model could be implemented as a function in a vehicle and verified using the *shadow mode approach*, that is, the output of this fully working function would be logged in the background without being actively used to trigger other functions or functionalities within an existing ADAS or AA system. Once the verification criteria have been satisfied during the series development stage, the system could be put live, adhering to one of the four *Automotive Safety Integrity Levels*, also known as *ASIL* (ISO 26262-9:2018, 2018).

In order to utilise human input effectively in AA systems, the human operator's, that is, driver's performance needs to be described formally and expressed numerically, in order to be used for any driver state estimation. This also applies to any semantic information contained in those driver state data. A possible approach to organise data to be processed in an AA context may lie in *ontologies*. Ontologies are formal, explicit specifications of a shared conceptualization (Gruber, 1993). Hereby, situations and states, requiring AA, could be evaluated and tracked, eventually leading to the development of an AA grammar as a subset of an all-encompassing traffic grammar, forming the foundation of sophisticated AA systems. Future work will seek to provide the empirical basis for such systems and for the CF within our driver model.

**Declaration of Competing Interest**

The authors declare that they have no known competing financial interests or personal relationships that could have appeared to influence the work reported in this paper.

**Acknowledgements**

The authors presented an initial version of the idea described in this paper at the 'Human Factors and Ergonomics Society Europe Chapter 2014 Annual Conference' held in Lisbon, Portugal in October 2014.

This research did not receive any specific grant from funding agencies in the public, commercial, or not-for-profit sectors.

**References**

Andrianov, A., Mohammadi Ziabari, S.S. & Gerritsen, C. (2021). A brain-inspired cognitive support model for stress reduction based on an adaptive network model. *Cognitive Systems Research, 65*, 151–166.   doi:10.1016/j.cogsys.2020.10.010

Baddeley, A. (2000). The episodic buffer: a new component of working memory? *Trends in Cognitive Sciences, 4*(11), 417–423.   doi:10.1016/S1364-6613(00)01538-2

Baddeley, A.D. & Hitch, G.J. (1994). Developments in the concept of working memory. *Neuropsychology, 8*, 485–493.

Balkenius, C. (1995). Natural Intelligence in Artificial Creatures. *Lund University Cognitive Studies, 37.*

Bingham, C.R. (2014). Driver Distraction: A Perennial but Preventable Public Health Threat to Adolescents. *Journal of Adolescent Health, 54*(5), S3–S5.   doi:10.1016/j.jadohealth.2014.02.015

Bubb, H. (2001). Haptik im Kraftfahrzeug. In T. Jürgensohn & K.-P. Timpe (Eds.), *Kraftfahrzeugführung* (pp. 155–175). Berlin: Springer-Verlag.

Burgess, P.W. & Simons, J.S. (2005). Theories of frontal lobe executive function: clinical applications. In P.W. Halligan & D.T. Wade (Eds.), *Effectiveness of Rehabilitation for Cognitive Deficits* (pp. 211–31). New York: Oxford University Press.

Collins, A. & Koechlin, E. (2012). Reasoning, Learning, and Creativity: Frontal Lobe Function and Human Decision-Making. *PLoS Biology 10*(3): e1001293.   doi:10.1371/journal.pbio.1001293

Diamond, A. (2013). Executive functions. *Annual Review of Psychology, 64*, 135–168. doi:10.1146/annurev-psych-113011-143750

Donges, E. (2009). Fahrerverhaltensmodelle. In H. Winner, S. Hakuli & G. Wolf (Eds.), *Handbuch Fahrerassistenzsysteme* (pp. 15–23). Wiesbaden: Vieweg+Teubner.

Fitts, P.M. (1951). *Human engineering for an effective air-navigation and traffic-control system*. Washington, DC: National Research Council.

Gavas, R.D., Tripathy, S.R., Chatterjee, D. & Sinha, A. (2018). Cognitive load and metacognitive confidence extraction from pupillary response. *Cognitive Systems Research, 52*, 325–334. doi:10.1016/j.cogsys.2018.07.021

Gruber, T.R. (1993). A translation approach to portable ontology specifications. *Knowledge Acquisition, 5*(2), 199–220. doi:10.1006/knac.1993.1008

Hull, C.L. (1952). *A Behavior System*. New Haven, CT: Yale University Press.

ISO 10075-1:2017 (2017). *Ergonomic principles related to mental workload - Part 1: General issues and concepts, terms and definitions (ISO 10075-1:2017)*. ISO. Retrieved January 04, 2021, from https://www.iso.org/standard/66900.html

ISO 26262-9:2018 (2018). *Road vehicles – Functional safety – Part 9: Automotive safety integrity level (ASIL)-oriented and safety-oriented analyses (ISO 26262-9:2018)*. ISO. Retrieved January 04, 2021, from https://www.iso.org/standard/68391.html

Kolb, B. & Whishaw, I.Q. (2009). *Fundamentals of Human Neuropsychology*. New York, NY: Worth Publishers.

Kramer, U. (2008). *Kraftfahrzeugführung*. München: Carl Hanser Verlag.

Lehto, J.E., Juujärvi, P., Kooistra, L. & Pulkkinen, L. (2003). Dimensions of executive functioning: evidence from children. *British Journal of Developmental Psychology, 21*, 59–80.

Lethaus, F. (2013). Reading the driver's eyes using machine learning. *Wessex Psychologist Bulletin, 9*, 20–22.

Lethaus, F. & Rataj, J. (2007). Do eye movements reflect driving manoeuvres? *IET Intelligent Transport Systems, 1*(3), 199–204. doi:10.1049/iet-its:20060058

Lethaus, F. & Rataj, J. (2008). Using eye movements as a reference to identify driving manoeuvres. In ATZ | ATZautotechnology (Ed.), *Proceedings of the FISITA World Automotive Congress 2008, vol. 1*. Wiesbaden: Springer Automotive Media.

Lethaus, F., Baumann, M.R.K., Köster, F. & Lemmer, K. (2011). Using pattern recognition to predict driver intent. In A. Dobnikar, U. Lotric & B. Šter (Eds.), *Adaptive and Natural Computing Algorithms, Lecture Notes in Computer Science, Part I, ICANNGA 2011* (vol. 6593, pp.140–149). Heidelberg: Springer. doi:10.5555/1997052.1997069

Lethaus, F., Baumann, M.R.K., Köster, F. & Lemmer, K. (2013). A comparison of selected simple supervised learning algorithms to predict driver intent based on gaze data. *Neurocomputing, 121*, 108–130. doi:10.1016/j.neucom.2013.04.035

Lethaus, F., Harris, R.M., Baumann, M.R.K., Köster, F. & Lemmer, K. (2013). Windows of driver gaze data: How early and how much for robust predictions of driver intent? In M. Tomassini, A. Antonioni, F. Daolio & P. Buesser (Eds.), *Adaptive and Natural Computing Algorithms, Lecture Notes in Computer Science, ICANNGA 2013* (vol. 7824, pp. 446–455). Heidelberg: Springer. doi:10.1007/978-3-642-37213-1_46

Lethaus, F., Sichler, B., Neukart, F., & Seidel, C. (2021). KI-Innovation über das autonome Fahren hinaus. In P. Buxmann & H. Schmidt (Eds.), *Künstliche Intelligenz: Mit Algorithmen zum* *wirtschaftlichen Erfolg* (pp. 81–100). Heidelberg: Springer. Retrieved April 20, 2021, from https://www.springer.com/de/book/9783662617939

Lunt, L., Bramham, J., Morris, R.G., Bullock, P.R., Selway, R.P., et al. (2012). Prefrontal cortex dysfunction and "jumping to conclusions": bias or deficit? *Journal of Neuropsychology, 6*, 65–78.

Meyer, D.E. & Schvaneveldt, R.W. (1971). Facilitation in recognizing pairs of words: Evidence of a dependence between retrieval operations. *Journal of Experimental Psychology, 90*, 227–234. doi:10.1037/h0031564

Meyer, D.E., Schvaneveldt, R.W. & Ruddy, M.G. (1975). Loci of contextual effects on visual word recognition. In P. Rabbitt & S. Dornic (Eds.), *Attention and Performance V* (pp. 98–118). London: Academic Press.

Mooneyham, B.W. & Schooler, J.W. (2013). The Costs and Benefits of Mind-Wandering: A Review. *Canadian Journal of Experimental Psychology, 67*(1), 11–18. doi:10.1037/a0031569

Posner, M.I. & DiGirolamo, G.J. (1998). Executive attention: conflict, target detection, and cognitive control. In R. Parasuraman (Ed.), *The Attentive Brain* (pp. 401–423). Cambridge, MA: MIT Press.

Rasmussen, J. & Rouse, W.B. (1981). *Human detection and diagnosis of system failures*. New York, NY: Plenum.

SAE International (2019). *SAE J3016: Levels of Driving Automation.* SAE International. Retrieved January 04, 2021, from https://www.sae.org/news/2019/01/sae-updates-j3016-automated-driving-graphic

Schießl, C. (2008a). Continuous subjective strain measurement. *IET Intelligent Transport Systems, 2*(2), 161–169. doi:10.1049/iet-its:20070028

Schießl, C. (2008b). Subjective strain estimation depending on driving manoeuvres and traffic situation. *IET Intelligent Transport Systems, 2*(4), 258–265. doi: 10.1049/iet-its:20080024

Schvaneveldt, R.W. & Meyer, D.E. (1973). Retrieval and comparison processes in semantic memory. In S. Kornblum (Ed.), *Attention and Performance IV* (pp. 395–409). New York, NY: Academic Press.

Sheridan, T.B. (2002). *Humans and automation: Systems design and research issues*. New York, NY: John Wiley.

Smallwood, J. & Schooler, J.W. (2009). Mind-wandering. In T. Bayne, A. Cleermans & P. Wilken (Eds.), *The Oxford Companion to Consciousness* (pp. 443–445). Oxford: Oxford University Press.

Smallwood, J. & Schooler, J.W. (2015). The Science of Mind Wandering: Empirically Navigating the Stream of Consciousness. *Annual Review of Psychology, 66*, 487–518. doi:10.1146/annurev-psych-010814-015331

Spence, K.W. (1956). *Behavior Theory and Conditioning*. New Haven, CT: Yale University Press.

Vasconez, J.P., Viscaino, M., Guevara, L. & Cheein, F.A. (2020). A fuzzy-based driver assistance system using human cognitive parameters and driving style information. Cognitive Systems Research, 64, 174–190. doi:10.1016/j.cogsys.2020.08.007

Vollrath, M. & Schießl, C. (2004). Belastung und Beanspruchung im Fahrzeug – Anforderungen an Fahrerassistenz. In VDI Wissensforum IWB GmbH (Ed.), *Integrierte Sicherheit und Fahrerassistenzsysteme* (pp. 343–360). Düsseldorf: VDI Verlag.

Wickens, C.D. & Kessel, C.J. (1979). The effects of participatory mode and task workload on the detection of dynamic system failures. *IEEE Transactions on Systems, Man and Cybernetics, 9*, 24–34. doi:10.1109/TSMC.1979.4310070

Wickens, C.D. & Kessel, C.J. (1980). Processing resource demands of failure detection in dynamic systems. *Journal of Experimental Psychology: Human Perception and Performance, 6*(3), 564–577. doi:10.1037/0096-1523.6.3.564

Wickens, C.D. & Kessel, C.J. (1981). Failure detection in dynamic systems. In J. Rasmussen & W.B. Rouse (Eds.), *Human detection and diagnosis of system failures* (pp. 155–169). New York, NY: Plenum.

Wickens, C.D., Mavor, A.S., Parasuraman, R. & McGee, P. (1998). Airspace system integration: The concept of free flight. In C.D. Wickens, A.S. Mavor & J.P. McGee (Eds.), *The future of air traffic control: Human operators and automation* (pp. 225–245). Washington, DC: National Academy.

Wiener, E.L. & Curry, R.E. (1980). Flight-deck automation: promises and problems. *Ergonomics, 23*(10), 995–1011.   doi:10.1080/00140138008924809

**Figures**

Figure 1 – Role of the Cognitive Factor within a driver model.

Figure 2 – Conceptual topology of the Cognitive Factor and multiplicatory relationships between cognitive sub-factor nodes.

Figure 3 – Example of a Cognitive Factor heatmap profile with an overall product value.

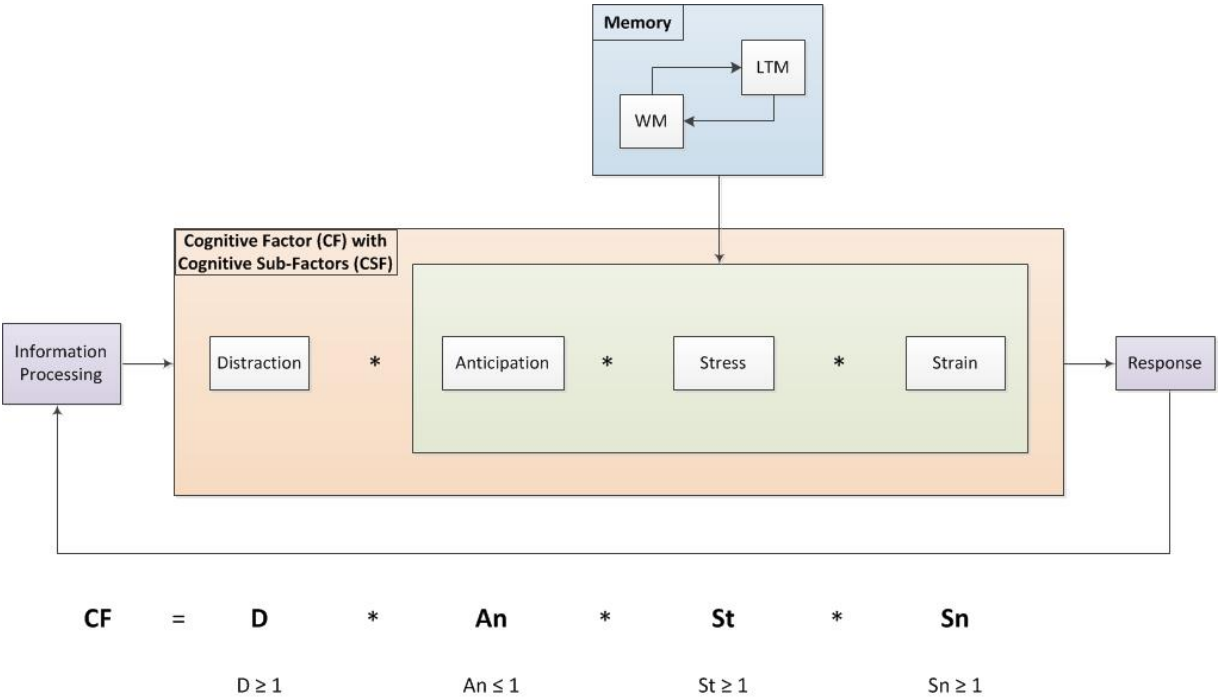

**Figure 1 – Role of the Cognitive Factor within a driver model.**

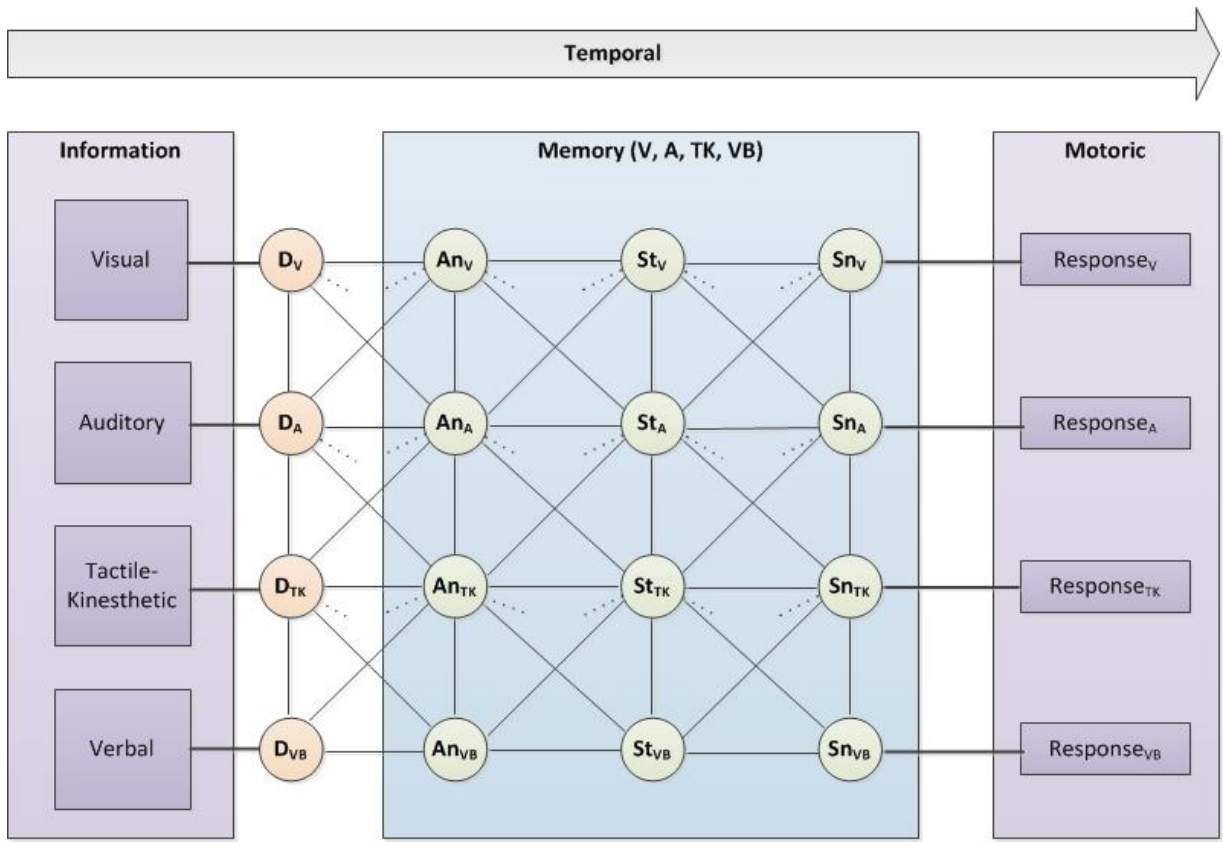

**Figure 2 – Conceptual topology of the Cognitive Factor and multiplicatory relationships between cognitive sub-factor nodes.**

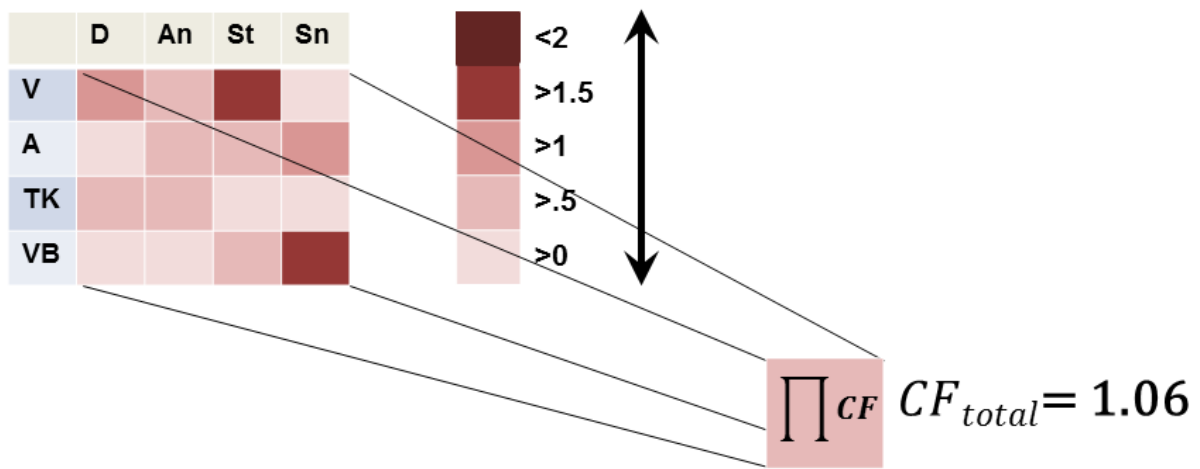

**Figure 3 – Example of a Cognitive Factor heatmap profile with an overall product value.**